\documentstyle[psfig,times]{mn}
\begin{document}
\hsize=6truein

\title[]{{\it Chandra} constraints on the thermal conduction in 
the intracluster plasma of A2142}

\author[]
{\parbox[]{6.in} {S. Ettori and A.C. Fabian\\
\footnotesize
Institute of Astronomy, Madingley Road, Cambridge CB3 0HA \\
}}                                            
\maketitle

\begin{abstract}
In this Letter, we use the recent {\it Chandra} observation of
Abell~2142 reported by Markevitch et al. to put constraints on thermal
conduction in the intracluster plasma. We show that the observed sharp
temperature gradient requires that classical conductivity has to be
reduced at least by a factor of between 250 and 2500. The result
provides a direct constraint on an important physical process relevant
to the gas in the cores of clusters of galaxies.
\end{abstract}


\section{INTRODUCTION} 
The heat stored in the intracluster plasma is conducted down any 
temperature gradient present in the gas in a way that can be described
through the following equations (Spitzer 1956, Sarazin 1988):
\begin{equation}
q =  \kappa \ \frac{d (kT_{\rm e})}{d r},
\label{eq:q}
\end{equation}
where $q$ is the heat flux, $T_{\rm e}$ is the electron
temperature, and $\kappa$ is the thermal conductivity
that can be expressed in term of the density, $n_{\rm e}$,
the electron mass, $m_{\rm e}$, and the electron mean free path,
$\lambda_{\rm e}$, as (Cowie and McKee 1977)
\begin{equation}
\kappa = 1.31 \ n_{\rm e} \ \lambda_{\rm e} \ \left(
\frac{kT_{\rm e}}{m_{\rm e}} \right)^{1/2}. 
\label{eq:k0}
\end{equation}

In a fully ionized gas of (mostly) hydrogen, the electron
mean free path is function of the gas temperature, density
and impact parameters of the Coulomb collisions, $\Lambda$:
\begin{equation}
\lambda_{\rm e}  =  30.2 \ {\sc T}^2 \ {\sl n}^{-1}
\ \left(\frac{\ln \Lambda}{37.9 + \ln \left(
{\sc T}/{\sl n}^{1/2} \right)} \right)^{-1} 
\mbox{kpc},
\label{eq:path}
\end{equation} 
where we have adopted the following dimensionless quantities:
\begin{equation}
{\sc T} = \left(\frac{kT_{\rm e}}{\mbox{10 keV}}\right),
\ {\sl n} = \left(\frac{n_{\rm e}}{\mbox{10$^{-3}$ cm$^{-3}$}}\right)
\end{equation}

Using this expression and the adopted typical values for cluster plasma, 
we can then write
\begin{equation}
\kappa = 8.2 \times 10^{20} \ {\sc T}^{5/2} \
\mbox{erg s$^{-1}$ cm$^{-1}$ keV$^{-1}$}.
\label{eq:k1}
\end{equation} 

If the mean electron free path is comparable to the scale length $\delta r$ of 
the temperature gradient, the heat flux tends to saturate to the 
limiting value which may be carried by the electrons (Cowie and McKee 1977):
\begin{eqnarray}
q_{\rm sat} & = & 0.42 \left( \frac{2 kT_{\rm e}}{\pi m_{\rm e}} \right)^{1/2}
n_{\rm e} \ kT_{\rm e} \nonumber \\
 & = & 0.023 \ {\sc T}^{3/2} \ {\sl n} \ \mbox{erg s$^{-1}$ cm$^{-2}$}, 
\label{eq:qsat}
\end{eqnarray}
where the factor 0.42 comes from the reduction effect on the heat conducted
by the electrons and that is produced from the secondary electric field 
that maintain the total electric current along the 
temperature gradient at zero (Spitzer 1956).

In this Letter, we apply these equations to estimate the efficiency
of thermal conductivity in the intracluster medium of A2142,
a cluster of galaxies observed by the X-ray telescope {\it Chandra}
during its calibration phase in August 1999. 

\section{THERMAL CONDUCTIVITY IN A2142}

Markevitch et al. (2000) have analyzed the {\it Chandra} observation
of the merging cluster of galaxies A2142 and made an important
discovery. The X-ray image reveals of sharp edges to the surface
brightness of the central elliptical-shaped region. The edges are
located about 3 arcmin to the northwest and 1 arcmin to the south with
respect to the X--ray centre. Markevitch et al (2000) show that the
bright and fainter regions either side of an edge are in pressure
equilibrium with each other, but with a dramatic electron temperature
decrease on the inside.

Considering the values of the intracluster gas properties in A2142
from their Figure~4, together with the equations presented in our
Introduction, we can estimate whether thermal conductivity is
efficient in erasing the observed temperature gradient.

The electron temperature (panel $b$ in their Fig.~4) varies from 5.8
to 10.6 keV on either side of the boundary of the southern edge of the
central bright patch in A2142, and from 7.5 to 13.8 keV at the
northern edge. The relative uncertainties on these values are about 20
per cent at the 90 per cent confidence level. The electron density
(panel $d$ in their Fig.~4) at the edges is $\sim$ 1.2 $\times$
10$^{-2}$ cm$^{-3}$, 3.0 $\times$ 10$^{-3}$ cm$^{-3}$ to the South and
North, respectively.

The scale length $\delta r$ on which this temperature gradient is
observed is spatially unresolved in the temperature profile and
appears enclosed between 0 and 35 kpc ($H_0 = $ 70 km s$^{-1}$
Mpc$^{-1}$) for the Southern edge, 0--70 kpc for the Northern edge.
However, the surface brightness profiles (panel $c$) show a radially
discontinuous derivative at the positions of the sharp edges, on
scales of about 10--15 kpc. We adopt hereafter these values as
representative of $\delta r$, using the larger values only for upper
limit purposes.

The case for a saturated flux is reached when these scales are
comparable with the electron mean free path of about 2 and 12 kpc for
the Southern and Northern edge, respectively, calculated using the
temperature and density estimates in eqn.~\ref{eq:path}.

Therefore, we will consider hereafter the two extreme cases where (i)
$\delta r \gg \lambda_{\rm e}$ and the heat flux is un-saturated, (ii)
$\delta r \approx \lambda_{\rm e}$ and the heat flux is saturated and
represented by eqn.~\ref{eq:qsat}.

The maximum heat flux in a plasma is given by
\begin{equation}
q = \frac{3}{2} \  n_{\rm e} \ kT_{\rm e} \ \bar{v},
\label{eq:flux}
\end{equation}
where $\bar{v} = d r / d \tau$ is a characteristic velocity that 
we are now able to constrain
equalizing the latter equation to eqn.~\ref{eq:q}.

In particular, given the observed values (and relative errors) of
density and temperature across the two edges and  $\delta r$ of (i:
non-saturated flux) 10 and 20 kpc (upper limits: 35 and 70 kpc for the
edges to South and North, respectively) and (ii: saturated flux) $\sim
\lambda_{\rm e}$, the characteristic time, $\delta \tau$, required to
erase the electron temperature gradient and due to the action of the
thermal conduction alone would be:
\begin{equation}
\delta \tau = \frac{\delta r}{\bar{v}} = \left\{ \begin{array}{l}
3.6 \ (<80) \ \times 10^6 \mbox{yrs}, \;  \delta r \gg 2 \mbox{kpc} \\
0.3 \ (<0.4) \ \times 10^6 \mbox{yrs}, \;  \delta r \approx 2 \mbox{kpc} 
\end{array}
\right.
\end{equation}
for the Southern edge, and 
\begin{equation}
\delta \tau = \frac{\delta r}{\bar{v}} = \left\{ \begin{array}{l}
2.4 \ (<52) \ \times 10^6 \mbox{yrs}, \;  \delta r \gg 12 \mbox{kpc} \\
1.9 \ (<2.4) \ \times 10^6 \mbox{yrs}, \;  \delta r \approx 12 \mbox{kpc}
\end{array}
\right.
\end{equation}
for the Northern edge.
The upper limits are obtained propagating the uncertainties on 
the temperature and, for the ``$\delta r \gg \lambda_{\rm e}$" 
condition only, assuming the spatial resolution of the temperature 
profile as indicative of the length of the gradient.
Here we note that the limit on the timescale for
saturated flux is the minimum value given the condition of the gas.
Any value of $\delta \tau$ estimated on scales considerably larger than
the electron mean free path has to be longer than the limit for 
saturated flux (also significantly, given that the timescale
is proportional to $\delta r^2$).

When this time interval is compared with the core 
crossing time of the interacting clumps of about 10$^9$ yrs, 
we conclude that thermal conduction needs to be suppressed by 
a factor larger than 10 and with a minimum characteristic 
value enclosed between 250 and 2500.
On the other side, Markevitch et al. suggest a dynamical model
in which the dense cores of the two interacting
clumps are moving through the host, less dense, intracluster medium
at a subsonic velocity of less than 1000 km s$^{-1}$ and
400 km s$^{-1}$, for the Northern and Southern
edge, respectively, leading to a timescale of
about $2 \times 10^7$ yrs or larger for the cool and hot phases
to be in contact. In this scenario, our results implies that the
conduction is suppressed {\it at least} by a factor of
2--200 in the South and 2--32 in the North.
However, we note that it is unlikely that the cooler gas can 
have arrived at its present arrangement and settled in the hotter
enviroment on a timescale much shorter than a core
crossing time. 
The frequency of the occurance of similar structures
in other cluster cores will be important in establishing
the timescale for their formation and duration and, hence, 
improve the constraint on the thermal conduction in the
intracluster plasma.

\section{CONCLUSIONS}
In this letter, we calculate the thermal conductivity in the intracluster
medium of A2142, a interacting cluster of galaxies observed by 
{\it Chandra} during the calibration phase.

We have shown that the time interval in which the action of thermal
conduction should propagate heat to neighbouring regions is shorter by
a factor of about 250--2500 than the likely estimated age of the
structure.
We note that the observed sharp temperature boundary
also means that mixing and diffusion are minimal.

The results presented here are a direct measurement of a physical
process in the intracluster plasma and imply that thermal conduction
is particularly inefficient within 280 $h_{70}^{-1}$ kpc of the
central core. The gas in the central regions of many clusters has a
cooling time lower than the overall age of the system, so that a slow
flow of hotter plasma moves here from the outer parts to maintain
hydrostatic equilibrium. In such a cooling flow (Fabian, Nulsen,
Canizares 1991; Fabian 1994), several phases of the gas (i.e. with
different temperatures and densities) are in equilibrium and would
thermalize if the conduction time were short. The large suppression of
plasma conductivity in the cluster core allows an inhomogeneous,
multi--phase cooling flow to form and be maintained, as is found from
spatial and spectral X-ray analyses of many clusters (e.g. Allen et
al. 2000).

How the conduction is reduced by a so large factor is still unclear.
Binney \& Cowie (1981) explain the reservoir for heat observed in the
region of M87 as requiring an rms field strength considerably larger
than the component of the field parallel to the direction along which
conduction occurs. (The transport processes are reduced in the
direction perpendicular to magnetic field lines). This would imply
either highly tangled magnetic fields or large--scale fields
perpendicular to the lines connecting the hotter to the cooler zones.
Such fields could become dynamically important.
Chandran, Cowley \& Albright (1999) use asymptotic analysis and 
Monte Carlo particle simulations to show that tangled field lines and,
with larger uncertainties, magnetic mirrors reduce the Spitzer
conductivity by a large factor.
Via a phenomenological approach Tribble (1989) argued that a multiphase
intracluster medium is an inevitable consequence of the effect of a
tangled magnetic field on the flow of the heat through the cluster
plasma. Electromagnetic instabilities driven by the 
temperature gradient (e.g. Pistinner, Levinson \& Eichler 1996) 
can represent another possible explanation for the suppression of
thermal conductivity.
Finally, we speculate that cooler gas dumped in the cluster core by a 
merger (see e.g. Fabian \& Daines 1991) would be part of a different 
magnetic structure to the hotter gas and so thermally isolated.

\section*{ACKNOWLEDGEMENTS} We acknowledge the support
of the Royal Society.

\end{document}